\PassOptionsToPackage{dvipsnames}{xcolor}
\documentclass[sigconf, anonymous=false]{acmart}
\usepackage{algorithm}
\usepackage{algorithmic}
\usepackage{bm}
\usepackage{makecell}
\usepackage{multirow}
\usepackage{subcaption}
\usepackage[capitalise,noabbrev]{cleveref}
\usepackage{svg}
\usepackage{xfrac}
\usepackage{xspace}
\usepackage{hyperref}

\newcommand{\allgatherv}{$\mathsf{Allgatherv}$\xspace}

\newcommand{\stdthreads}{$\mathsf{std::threads}$\xspace}
\newcommand{\cudamemcpy}{$\mathsf{cudamemcpy}$\xspace}

\newcommand{\B}{\mathbf{B}\xspace}
\newcommand{\A}{\mathbf{A}\xspace}
\newcommand{\C}{\mathbf{C}\xspace}
\newcommand{\li}{\textsc{Li}\xspace}
\newcommand{\gi}{\textsc{Gi}\xspace}
\newcommand{\kokkos}{\textsc{KokkosKernels}\xspace}

\newcommand{\name}{\textsc{Trident}\xspace}
\newcommand{\ours}{\name}
\newcommand{\spgemm}{SpGEMM\xspace}
\newcommand{\summa}{Improved Sparse SUMMA\xspace}

\newcommand{\tridentpart}{trident partitioning\xspace}
\newcommand{\totalnodes}{N}
\newcommand{\distspgemmalgo}{distributed SpGEMM algorithm\xspace}

\newcommand{\gridedge}{\sqrt{\frac{P}{\lambda}}}
\newcommand{\gridedgecompact}{\sqrt{P/\lambda}}
\newcommand{\coarsergridcompact}{\gridedgecompact\times\gridedgecompact}
\newcommand{\tridentgrid}{\gridedge \times \gridedge \times \lambda}
\newcommand{\tridentgridcompact}{\gridedgecompact \times \gridedgecompact \times \lambda}

\title{Communication-Avoiding SpGEMM via Trident Partitioning on Hierarchical GPU Interconnects
} 

\date{December 2025}

\begin{abstract}
The multiplication of two sparse matrices, known as \spgemm, is a key kernel in scientific computing and large-scale data analytics, underpinning graph algorithms, machine learning, simulations, and computational biology, where sparsity is often highly unstructured.

The unstructured sparsity makes achieving high performance challenging because it limits both memory efficiency and scalability. 
In distributed memory, the cost of exchanging and merging partial products across nodes further constrains performance.
These issues are exacerbated on modern heterogeneous supercomputers with deep, hierarchical GPU interconnects. 
Current \spgemm implementations overlook the gap between intra-node and inter-node bandwidth, resulting in unnecessary data movement and synchronization not fully exploiting the fast intra-node interconnect.

To address these challenges, we introduce \name, a hierarchy-aware 2D distributed \spgemm algorithm that uses communication-avoiding techniques and asynchronous communication to exploit the hierarchical and heterogeneous architecture of modern supercomputing interconnect. 
Central to \name is the novel \textbf{trident partitioning} scheme, which enables hierarchy-aware decomposition and reduces internode communication by leveraging the higher bandwidth between GPUs within a node compared to across nodes.
Here, we evaluate \name on unstructured matrices, achieving up to {$2.38\times$} speedup over a 2D \spgemm with a corresponding geometric mean speedup of $1.54\times$. 
\name reduces internode communication volume by up to $2\times$ on NERSC's Perlmutter supercomputer. 
Furthermore, we demonstrate the effectiveness of \name in speeding up Markov Clustering, achieving up to $2\times$ speedup compared to competing strategies.
\end{abstract}

\copyrightyear{2026}
\acmYear{2026}
\setcopyright{cc}
\setcctype{by}
\acmConference[ICS '26]{2026 International Conference on Supercomputing}{July 06--09, 2026}{Belfast, United Kingdom}
\acmBooktitle{2026 International Conference on Supercomputing (ICS '26), July 06--09, 2026, Belfast, United Kingdom}
\acmDOI{10.1145/3797905.3800543}
\acmISBN{979-8-4007-2522-7/2026/07}

\begin{document}

\author{Julian Bellavita}
    \authornote{Both authors contributed equally to this research.}
    \affiliation{%
    \institution{Cornell University}
    \city{Ithaca, NY}
    \country{USA}}
\email{jbellavita@cs.cornell.edu}

\author{Lorenzo Pichetti}
    \authornotemark[1]
    \affiliation{%
    \institution{University of Trento}
    \city{Trento}
    \country{Italy}
}
\email{lorenzo.pichetti@unitn.it}

\author{Thomas Pasquali}
\affiliation{%
  \institution{University of Trento}
  \city{Trento}
  \country{Italy}
}
\email{thomas.pasquali@unitn.it}

\author{Flavio Vella}
\affiliation{%
  \institution{University of Trento}
  \city{Trento}
  \country{Italy}
}
\email{flavio.vella@unitn.it}

\author{Giulia Guidi}
\affiliation{%
  \institution{Cornell University}
  \city{Ithaca, NY}
  \country{USA}
}
\email{gguidi@cornell.edu}

\renewcommand{\shortauthors}{Bellavita et al.}

\maketitle

\section{Introduction}

General sparse matrix multiplication (\spgemm) is a fundamental building block in many scientific areas~\cite{bulucc2025ubiquitous}, including graph algorithms~\cite{azad15, davis2019algorithm}, computational biology~\cite{guidi2021parallel, guidi2021bella, guidi2022distributed, selvitopi2020distributed, besta2020communication}, clustering~\cite{hipmcl}, and iterative solvers such as algebraic multigrid (AMG)~\cite{dalton2015data, vanek1994algebraic}. 
In these cases, sparsity often arises from irregular connectivity, multilevel coarsening, or permutation rather than from an underlying geometric stencil. As a result, nonzeros are typically uniformly distributed, producing sparse matrices that are naturally load balanced; we refer to such matrices as \textit{unstructured}.
This class of matrices differs from structured matrices commonly found, for example, in PDE discretizations, where strong geometric locality creates a patterned nonzero distribution, such as banded or stencil-based patterns which can be directly exploited for communication optimization. 
Unstructured matrices are often load balanced, but their lack of predictable nonzero structure complicates communication optimization on modern supercomputing systems.

In protein similarity searches, the sparsity pattern is influenced by the distribution of shared subsequences between sequences \cite{selvitopi2020distributed}, while genome assembly similarly produces an irregular pattern due to technological noise and genomic repetition~\cite{guidi2021bella, guidi2021parallel}.
Markov Clustering (MCL)~\cite{van2008graph} also produces unstructured matrices.
Both the random-walk expansion and the pruning step remove any consistent locality, and the input is often unstructured as well~\cite{hipmcl, li2022fast}.

In practice, structured sparse matrices are often found in regular linear systems produced by discretization techniques such as the finite element method~\cite{janna2009comparison} or in some optimization problems~\cite{schenk2009inertia}.
These problems can sometimes bypass SpGEMM, instead leveraging the underlying structure to efficiently reach a solution using dense primitives~\cite{wang2024optimization}.
If SpGEMM is needed~\cite{hong24}, 1D partitioning strategies, where each process owns a block of rows or columns of the matrix, are often preferred because they reduce load imbalance across processes for structured matrices.
In this paper, we focus on \textit{unstructured} matrices due to their widespread presence in both traditional and emerging large-scale sciences.

The dominant 2D formulation for distributed SpGEMM, introduced by Bulu\c{c} and Gilbert~\cite{buluc2008challenges,buluc11b}, divides the matrices over a two-dimensional process grid with an equal number of processes along each dimension. 
As a result, each process owns only a subset of rows and a subset of columns, rather than an entire block row or block column. 
Both 1D and 2D formulations can balance nonzeros for unstructured matrices; however, 2D approaches require substantially less communication and avoid sparsity-aware communication, making them more efficient for large-scale SpGEMM on unstructured problems.
Current state-of-the-art 2D approaches include distributed Sparse SUMMA~\cite{buluc11b}, as implemented in the CombBLAS library~\cite{comblass}, which relies on bulk-synchronous collective communication and the RDMA-based Sparse SUMMA proposed by Brock et al.~\cite{brock2024rdma}. 
Both provide GPU support, but neither is designed for modern deep hierarchical interconnect, where the cost of intranode and internode communication can differ significantly.

In general, SpGEMM performance at scale is heavily influenced by data movement, as the cost of exchanging nonzeros can exceed the cost of arithmetic computation.
Today's large-scale machines use hierarchical interconnect that creates a significant bandwidth gap between intra- and internode communication. 
GPUs within a node connect via high-bandwidth, low-latency interconnect such as NVLink or Infinity Fabric, providing intranode bandwidth of around 900 GB/s per GPU~\cite{nvidia_nvlink_datacenter_2025}.
Communication across nodes uses higher-latency interconnect such as InfiniBand or HPE Cray Slingshot~\cite{hpe_cray_supercomputing_2025}, with typical bandwidth of around 400 Gb/s per port.
As the GPUs per node increase and CPU-GPU integration improves, the bandwidth gap between intranode and internode interconnect is expected to persist or widen~\cite{fusco2024understanding, tarraga2025understanding}. 
Current SpGEMM approaches do not consider this hierarchical network, leaving opportunities for optimized communication schedules untapped.

In this work, we present \name, a hierarchy-aware distributed \spgemm algorithm that uses communication-avoiding techniques and asynchronous messages to leverage hierarchical, heterogeneous interconnect. 
\ours is named for the novel \textbf{trident partitioning} scheme presented in this paper, which uses 2D partitioning between nodes and 1D partitioning within a node.
\name reduces costly internode communication by structuring the entire communication schedule into two phases. 
It first establishes internode data exchanges using peer-to-peer transfers, then uses the high-bandwidth intranode interconnect to stage, aggregate, and reuse data locally .
This design enables efficient pipelining of internode transfers with intranode collectives, hiding latency and minimizing internode communication.
In particular, our trident partitioning scheme and asynchronous progress engine are designed to (i) maximize GPU–GPU locality within a node, (ii) minimize volume and contention on the Slingshot/InfiniBand interconnect, and (iii) enable overlap between communication and computation.

Our results show that \name is consistently faster than the 2D approach, with a speedup of up to $2.38\times$ and a geomean speedup of $1.54\times$ across a set of large test matrices at 256 GPUs.
\name achieves up to $5.95\times$ speedup and a geomean speedup of $2.96\times$ over the sparsity-aware 1D SpGEMM algorithm in Trilinos. 
Our hierarchical communication strategy reduces internode communication by up to $2\times$. 
Finally, when used to implement Markov Clustering, \name provides up to a $2\times$ speedup over a version of Markov Clustering implemented with our fastest baseline.
\vspace{-.5em}
\section{Background and Related Work}\label{sec:background}

Given sparse matrices $\A \in \mathbb{R}^{m \times k}$ and $\B \in \mathbb{R}^{k \times n}$, sparse matrix--matrix multiplication computes $\C = \A\B$, where $\C \in \mathbb{R}^{m \times n}$. 
In \name, matrices are stored in the compressed sparse row (CSR) format, which encodes each matrix using three arrays: a row-pointer array \texttt{rowptr} of length $m+1$, a column-index array \texttt{colind} of length \texttt{nnz} (i.e., number of nonzeros), and a value array \texttt{val} also of length \texttt{nnz}.
CSR enables efficient row access and merging of column indices, making it popular for SpGEMM on both CPUs and GPUs. 
For the local multiplication on GPU, we use \kokkos~\cite{rajamanickam2021kk, DEVECI201833}.

\subsection{Distributed SpGEMM}

In this section, we review recent work on distributed SpGEMM and highlight key differences from our proposed approach. 
The algorithms are categorized based on how the matrices are partitioned across the processes.

\subsubsection{1D and Sparsity-Aware Distributed SpGEMM}

Bulu\c{c} et al.~\cite{buluc2008challenges} first formalized a sparsity-agnostic 1D algorithm. 
Their 1D algorithm divides $\A$ in a block-row fashion and replicates $\B$ on every process. 
1D approaches are generally simpler and effective for structured matrices, but the communication increases for unstructured matrices and a large number of processes~\cite{boman13,ballard2016hypergraph}.
To reduce unnecessary communication, sparsity-aware 1D strategies have been proposed. 
Ballard et al.~\cite{ballard13} showed that for Erd\H{o}s--R\'enyi matrices with sparsity ratio $d$, 1D sparsity-aware strategies can outperform 2D sparsity-agnostic approaches by a factor of $\mathcal{O}(\sqrt{P}/d)$ in communication volume. 
However, CPU implementations faced significant challenges due to message packing and the irregular data access required by sparsity awareness~\cite{azad15,buluc11b}.
To address these challenges, Hong and Bulu\c{c}~\cite{hong24} introduced an RDMA-based 1D sparsity-aware scheme that hides packing overhead using one-sided communication; however, their approach requires communicating a whole block even when only a single row is needed.
Ranawaka et al.~\cite{ranawaka2024distributed} presented a 1D sparsity-aware SpGEMM algorithm optimized for a tall-and-skinny right-hand operand. 
These approaches are limited to CPUs.
PETSc~\cite{petsc} and Trilinos~\cite{trilinos} use distributed Gustavson-style algorithms that partition the rows of $\A$ and exchange the necessary rows of $\B$, implementing sparsity-aware communication.
Trilinos uses \kokkos for GPU acceleration, while PETSc provides CUDA and HIP backend. 

To our knowledge, the closest hierarchical approach is Zhang et al.~\cite{zhang2025cola}, who apply hierarchy-aware communication to 1D partitioned SpMM. 
However, SpMM is fundamentally different from SpGEMM because it involves a dense operand.
Zhang et al. use reordering to induce a near-diagonal sparsity structure, an assumption that does not hold for the unstructured matrices targeted in this work, where 2D partitioning is often more effective.
The Trident partitioning proposed in this work creates a qualitatively different communication pattern compared to 1D approaches and does not depend on structural information about the input matrices.

\subsubsection{2D Distributed SpGEMM}

The Sparse SUMMA algorithm by Bulu\c{c} and Gilbert~\cite{buluc2008challenges} is the dominant 2D formulation and significantly reduces communication compared to 1D approaches, especially for unstructured and irregular matrices. 
The algorithm is implemented in the CombBLAS library~\cite{comblass}. 
This method distributes the matrices over a $\sqrt{P} \times \sqrt{P}$ process grid. Consequently, each process owns only a subset of rows and a subset of columns, rather than an entire block row or block column. 
The algorithm proceeds in $\sqrt{P}$ stages; at each stage, a submatrix of $\A$ is broadcast along a processor row and a submatrix of $\B$ is broadcast along a processor column, resulting in a per-processor communication cost proportional to $\mathcal{O}(\sqrt{P})$. 
Each processor participates in \(\mathcal{O}(\sqrt{P})\) collective communication stages, incurring a latency cost of \(\mathcal{O}(\sqrt{P})\) while simultaneously reducing per-processor communication volume by a factor of \(\sqrt{P}\) compared to 1D schemes, improving bandwidth efficiency.
Under bulk-synchronous parallel (BSP) cost model, this is communication-optimal up to a logarithmic factor. 
The reduced volume enhances scalability at large process counts, but the $\sqrt{P}$ synchronization stages add non-negligible latency, which can limit performance on modern large-scale systems.

Cannon's algorithm operates on a 2D process grid, beginning with a \textit{stagger} step that circularly shifts submatrices of $\A$ and $\B$ to align matching pairs for the first local multiplication~\cite{cannon1969cellular}. 
Each process shifts its portion of $\A$ left by its row index and $\B$ up by its column index. 
In each iteration $r$, processes compute a partial product for $\C$ and then shift $\A$ left and $\B$ up along their respective rows and columns. 
Once $P$ iterations are complete, the partial products are accumulated to form $\C_{ij}$.
In the sparse variant~\cite{bulucc2010highly}, hypersparse kernels are used for local multiplication, and only nonempty submatrices are communicated, reducing both arithmetic and bandwidth while preserving the 2D shift schedule.

Recent work has explored 2D SpGEMM on GPU systems. 
Brock et al.~\cite{brock2024rdma} introduced a GPU-based Sparse SUMMA implementation that uses NVSHMEM-based one-sided RDMA to mitigate load imbalance and reduce synchronization overhead~\cite{nvshmem_lib}.
McFarland et al.~\cite{mcfarland2025parallel} proposed a hybrid host- and device-initiated communication strategy that dynamically switches between the CPU and GPU based on message size to reduce latency and improve throughput. 
Their implementation is based on CombBLAS, and their use of the GALATIC local SpGEMM kernel enables semiring abstraction~\cite{Lett_2021}. 

\subsubsection{3D Distributed SpGEMM}

3D SpGEMM approaches assign subcubes of $\A$, $\B$, and $\C$ to processes arranged on a $p = p_r \times p_c \times p_\ell$ grid indexed by $P_{ijk}$, and exchange entries of $\A$ and $\B$ as well as partial sums of the intermediate product of $\C$.
Classical 3D schemes replicate input or output matrices. 
In contrast, Azad et al.~\cite{azad2016exploiting} propose an iterative 3D SpGEMM that distributes $\A$, $\B$, and $\C$ over a $p_r \times p_c \times p_\ell$ process grid without explicitly replicating input or output entries.
In the simplified configuration $p_r = p_c = \sqrt{\sfrac{P}{p_\ell}}$ and $p_\ell = c$, each 2D layer behaves like a $\sqrt{\sfrac{P}{c}} \times \sqrt{\sfrac{P}{c}}$ grid, so splitting submatrices along the third dimension reduces communication by a factor of $\sqrt{c}$ while incurring only sparsity-dependent extra memory for the intermediate product. 
By replicating submatrices along the third dimension, the algorithm trades increased memory usage for fewer messages and lower communication volume.
Hussain et al.~\cite{hussain2021communication} developed a related approach targeting extremely large output matrices under a tight memory constraint.
They are designed for distributed-memory CPU machines and do not support GPUs. 

\subsection{Hierarchical Network}\label{sec:background:interconnect}

SpGEMM performance at large scale is dominated by data movement rather than computation. 
GPUs within a node are connected by high-bandwidth, low-latency interconnects such as NVLink or Infinity Fabric~\cite{nvidiaNVLinkNVSwitch, pearson2023interconnect}, but distributed SpGEMM algorithms largely fail to exploit this hierarchy. 
They typically treat intranode and internode communication uniformly, underutilizing the fast local interconnect and increasing latency and network traffic, which limits the performance benefits of modern node architectures.

If communication crosses multiple nodes, it relies on comparatively slower, higher-latency, and lower-bandwidth interconnects such as InfiniBand or HPE Cray Slingshot. 
De Sensi et al.~\cite{10793179} showed that intranode bandwidth is up to 8$\times$ higher than internode interconnect bandwidth. 
For example, NERSC's Perlmutter nodes have four NVIDIA A100 GPUs connected via NVLink and scale across nodes using the Slingshot-11 dragonfly network~\cite{nerscPerlmutterArch}. 
Each node has multiple Slingshot NICs because internode bandwidth is limited.
DGX and HGX servers also provide high intranode GPU bandwidth ($\approx$ 900 GB/s per GPU and multi-TB/s bisection), but internode communication relies on InfiniBand NDR, which provides 400 Gb/s per port~\cite{banchelli2024nvidia}.
Recent work has shown that adding more GPUs per node increases NIC pressure, exacerbating the imbalance~\cite{tarraga2025understanding}. Overall, the gap between intra- and internode bandwidth is expected to persist or widen as hardware advances~\cite{fusco2024understanding}.

\subsubsection{Hierarchy-Aware Graph Computation}

Pichetti et al. exploit hierarchical GPU network by combining 3D graph partitioning with communication-aware breadth-first search and amortized global synchronization to enable scalable approximate betweenness centrality on large multi-GPU systems~\cite{pichetti2025macbeth}.
GraphCube also focuses on hierarchical HPC network, co-optimizing topology-aware graph partitioning and routing-table configuration to balance communication across the network hierarchy~\cite{gan2024graphcube}.

\subsubsection{Hierarchy-Aware Collective Communication}

Early MPI work and recent GPU libraries such as HiCCL show that leveraging intranode and internode structure can improve the performance of standard collective communication~\cite{zhu2009hierarchical, hidayetoglu2025hiccl, temuccin2022accelerating}. 
HiCCL provides a composable library of hierarchical collectives optimized for NVLink, Infinity Fabric, PCIe, and internode network, with an emphasis on performance portability~\cite{hidayetoglu2025hiccl}.
TACCL optimizes collective communication on heterogeneous GPU interconnect by explicitly modeling link cost and routing, and by synthesizing machine-specific collectives from user-defined communication sketches~\cite{shah2025taccl}.
However, TACCL's scalability is limited by the cost of code synthesis based on an SMT solver.
These approaches focus on reusable collective primitives and generally assume communication patterns that are static, regular, and known a priori.

\begin{table}[t]
\caption{A set of notations used in this paper.}
\label{tab:symbols}
\centering
\resizebox{.95\linewidth}{!}{%
\begin{tabular}{@{}ll@{}}
\toprule
\textbf{Symbol} & \textbf{Definition} \\
\midrule
$P$ & The number of total MPI processes \\
$P_{ij}$  & A given process or GPU on a 2D grid\\
$P_{ijk}$ & A given process or GPU on a 2D-1D hybrid grid\\
$P_{ij:}$ & ``:'' indicates any third index on a 2D-1D hybrid grid \\ 
$\totalnodes$ & The total number of nodes \\
$\totalnodes_{ij}$ & A given node on the 2D processor grid \\
\li & The local interconnect \\
\gi & The global interconnect \\
$\lambda$ & The number of GPUs per node connected via \li \\
$r$ & Current local SpGEMM round \\
$\A$, $\B$, $\C$ & The left operand, right operand, and result of $\C=\A\B$ \\
$\B_{ij}$ & 2D tile of $\B$ on a 2D-1D hybrid grid \\
$\A_{ijk}$, $\B_{ijk}$, $\C_{ijk}$ & $\A$, $\B$, or $\C$ tile on ae 2D-1D hybrid grid \\
\bottomrule
\end{tabular}\label{tab:symbols}
}
\end{table}

In contrast, SpGEMM in scientific computing and biology generates sparsity-driven, data-dependent communication patterns that vary across inputs and iterations, making them difficult to express with static collectives or communication sketches. 
\name uses NCCL collectives for efficient local communication within a node, while non-blocking point-to-point messages are used for irregular internode communication.
For an overview of GPU-centric communication libraries and the use of hierarchical interconnect, we refer to the survey by Unat et al.~\cite{unat2024landscape}.

In summary, the literature indicates that although 1D sparsity-aware techniques effectively reduce communication for structured matrices, they provide limited benefits for unstructured matrices and scale poorly beyond moderate process counts~\cite{buluc2008challenges, hong24, boman13, ballard2016hypergraph, demirci2020cartesian}.
In contrast, 2D approaches communicate less and scale better for unstructured sparse matrices.
Critically, prior SpGEMM approaches assume a uniform interconnect and do not account for hierarchical network differences between intranode and internode bandwidth on modern GPU machines.
This gap motivates \name, a hierarchy-aware distributed GPU-based SpGEMM algorithm based on a new trident partitioning strategy designed to exploit hierarchical network and prioritize faster intranode communication.
\section{Algorithm}\label{sec:algorithm}

In this section, we introduce \name, a hierarchy-aware distributed SpGEMM algorithm designed to account for the characteristics of the interconnect. 
Rather than treating communication uniformly, \name prioritizes fast intranode communication and regulates slow inter-node data movement. 
This is accomplished by rethinking both the partitioning of the matrices and the organization of the computation, ensuring that the algorithm's communication paths align with the hardware's bandwidth hierarchy.
The core idea behind \name is a hybrid 2D–1D partitioning, or trident partitioning, strategy that assigns coarse-grained 2D matrix tiles to entire compute nodes, while further subdividing each tile into fine-grained 1D slices distributed across the GPUs within a node.
Unlike 3D SpGEMM, the additional dimension is not used to replicate submatrices but to reorganize communication so that inter-node transfers occur only once per node, followed by aggregation over the fast local interconnect. 
\name moves away from bulk-synchronous approaches by employing an asynchronous, $\C$-stationary execution model, allowing independent output tiles to progress without global synchronization. 
This overlap of global communication, computation, and intranode data movement reduces idle time and improves scalability for unstructured sparse matrices.

In this section, we describe our novel \tridentpart and the \name algorithm derived from it. First, we define a formal model for hierarchical interconnect that captures the fundamental asymmetry between intranode and inter-node communication. 
Using this model, we then present our 2D-1D hybrid \tridentpart and describe the algorithmic choices that align its communication schemes with the bandwidth hierarchy of modern GPU systems.
Table~\ref{tab:symbols} summarizes useful notation used throughout the paper.

\begin{figure}[t]
    \centering
    \includegraphics[width=\linewidth]{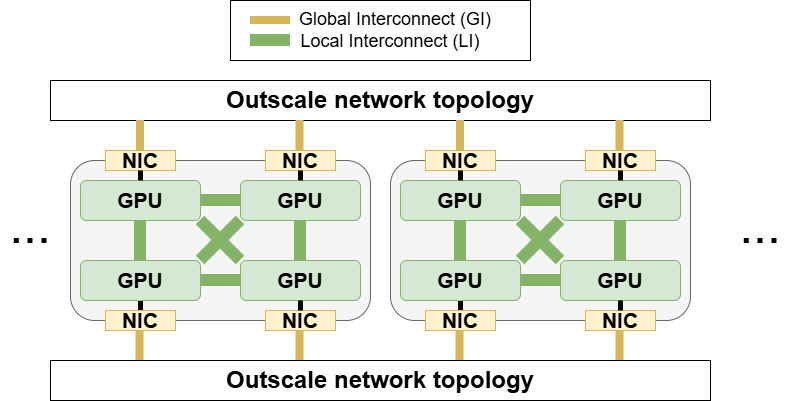}
    \caption{
    A model of a hierarchical interconnect and compute subsystem, where each node contains four GPUs connected internally via \li\ and across nodes via \gi.
    }
    \label{fig:networkexample}
\end{figure}

\begin{figure*}[t]
    \centering
    \begin{subfigure}[b]{0.48\textwidth}
        \centering
        \includegraphics[width=\textwidth]{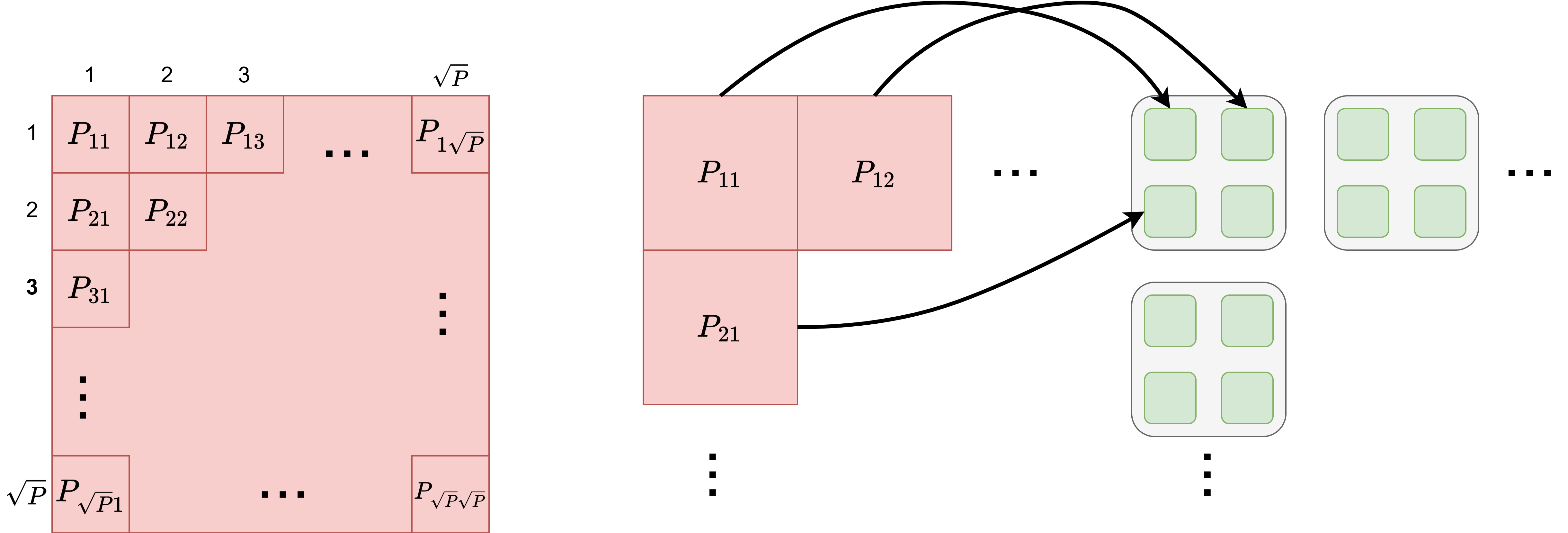}
        \caption{Classic 2D Partitioning}
        \label{sfig:partitioning:std}
    \end{subfigure}
    \hfill
    \begin{subfigure}[b]{0.48\textwidth}
        \centering
        \includegraphics[width=\textwidth]{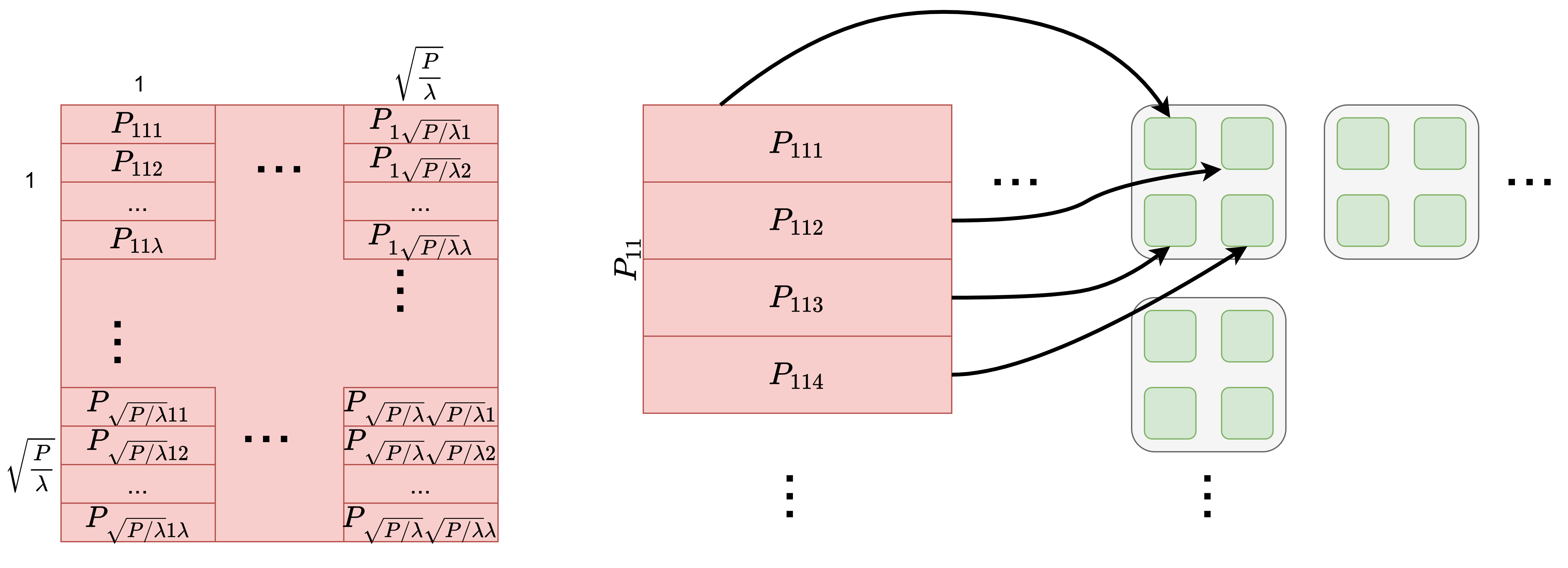}
        \caption{\name Partitioning}
        \label{sfig:partitioning:our}
    \end{subfigure}
    \caption{Each process is mapped to a GPU on a node. (a) In Sparsa SUMMA, processes are organized in a $\sqrt{P} \times \sqrt{P}$ grid. (b) In \name, processes are organized using a hybrid 2D $+$ 1D scheme, resulting in a $\tridentgridcompact$ grid.}
    \label{fig:partitioning}
\end{figure*}

\subsection{Hierarchical Network Definition}

Consider a distributed multi-GPU system with multiple compute nodes, each including several GPUs connected by a high-bandwidth, low-latency intranode interconnect, and linked to other nodes through a lower-bandwidth, higher-latency network. Such systems are typical of current leadership-class supercomputers, where GPU-to-GPU communication within a node is often significantly faster than communication between nodes.
Thus, we model the communication substrate as a two-level hierarchical interconnect. 
The global interconnect, \gi, enables communication between GPUs on different nodes, while the local interconnect, \li, connects GPUs within the same node. 
The global interconnect \gi generalizes technologies such as InfiniBand and HPE Cray Slingshot, while the local interconnect \li generalizes high-bandwidth GPU fabrics such as NVLink and Infinity Fabric.
Communication over \li has substantially lower latency and higher sustainable bandwidth than over \gi, and this performance gap persists as system scale increases. 

\cref{fig:networkexample} illustrates systems where GPUs within the same node are connected through \li, while GPUs on different nodes are connected through \gi.
$P$ denotes the total number of MPI processes, with exactly one process per GPU. 
GPUs are grouped into nodes of size $\lambda$, representing the number of GPUs connected via the same local interconnect, resulting in $N = P / \lambda$ nodes. 
Communication between processes within the same node occurs exclusively over \li, while communication between processes on different nodes must traverse \gi. 
This abstraction does not assume uniformity within either interconnect; variations in link topology, routing, or contention are incorporated into the effective bandwidth and latency parameters of \li and \gi.
In our experiment, we use NERSC's Perlmutter supercomputer, which has 4 GPUs per node; that is, $\lambda = 4$.
In addition, we assume $P$ is a square number (i.e., $\sqrt{P} \in \mathbf{N}$).

\subsection{\name Partitioning}

The classic 2D approach divides the input matrix into a $\sqrt{P} \times \sqrt{P}$ grid, as shown in \cref{sfig:partitioning:std}. 
The input matrix is split into tiles (red squares), and each tile is assigned to a process or GPU (green squares).
The underlying partitioning and tile assignment are oblivious to the network hierarchy.
In contrast, our \tridentpart scheme is designed to align with this hierarchy. 
By making network awareness a first-class algorithmic principle, \tridentpart enables a fundamentally more efficient use of the local interconnect. 

The first level of the trident partitioning scheme uses a 2D decomposition. The input matrix is divided into 2D tiles, which are organized into a coarser-grained grid of size $\coarsergridcompact$. 
This coarse grid forms the top tier of \name's hierarchy, grouping tiles into larger regions than traditional 2D approaches such as Sparse SUMMA, to reduce global communication and improve scalability.
Once this coarse-grained 2D partitioning is established, each tile within the coarse grid is further subdivided into $\lambda$ one-dimensional slices, which are distributed among individual processes (i.e., GPUs). 
This approach creates a hybrid 2D $+$ 1D partitioning scheme, resulting in a three-dimensional process grid $\tridentgridcompact$. 
The first two dimensions correspond to the coarse 2D tiles, and the third dimension indexes the 1D slices within each tile.
The slices derived from the same 2D tile are assigned to processes within the same \li group, ensuring that any pair of processes $P_{ijk}$ and $P_{ijh}$ can communicate via the local interconnect. In contrast, communication between processes $P_{ijk}$ and $P_{i'j'k}$, where $i \neq i'$ or $j \neq j'$, must traverse the global interconnect.
\cref{sfig:partitioning:our} illustrates the resulting partitioning and tile assignment in \name, highlighting the differences compared to the traditional 2D approach.
The presented hierarchical partitioning strategy is implemented in \name to minimize costly global communication while leveraging high-bandwidth, low-latency interconnect within nodes. 

Our partitioning uses three-dimensional indices but differs fundamentally from traditional 3D partitioning.
The third dimension does not represent subcubes but instead provides a finer-grained repartitioning of the coarser 2D grid into 1D slices. To avoid confusion, we refer to our scheme as a hybrid 2D $+$ 1D partitioning or \tridentpart rather than a 3D partitioning.

\subsection{\name Algorithm}

Given the trident partitioning, we now present the \name algorithm, first describing its 2D component and then its 1D component.

\subsubsection{\name Outer 2D Algorithm}\label{sec:algorithm:outeralg}

In our partitioning, 2D tiles are assigned to entire computation nodes, each composed of $\lambda$ processes, as shown in \cref{sfig:partitioning:our}, rather than to individual processes.
Due to the advantages of using RDMA demonstrated by Brock et al.~\cite{brock2024rdma}, \name uses an asynchronous $\C$-stationary \distspgemmalgo, in which each 2D tile can compute independently without global synchronization. 
In a $\C$-stationary scheme, the output matrix $\C$ is partitioned among processes, and each process is responsible for a fixed subset of $\C$ throughout the computation. 
This ensures that updates to each block are local, eliminating the need for global coordination. 
Each tile can proceed as soon as its input dependencies are available.
Common communication primitives, including collectives and even non-blocking MPI Isend/Irecv, cannot achieve this level of asynchrony because they require either a synchronization barrier or prior knowledge of other processes' rank information. 
The key difference between \name and the traditional 2D is the data access pattern.

In traditional 2D schemes, multiplication is organized as a sequence of iterations over block panels of the inner dimension. 
For each iteration $r$, each process $P_{ij}$ computes the partial product:
\begin{equation}\label{eq:2daccess}
    \C_{ij}^{(r)} = \A_{ir} \cdot \B_{rj},
\end{equation}
\noindent
$\A_{ir}$ and $\B_{rj}$ are sparse submatrices from the $r$-th block column of $\A$ and the $r$-th block row of $\B$, respectively. 
The final block $\C_{ij}$ is obtained by accumulating Eq.~\ref{eq:2daccess} over $r$ iterations.

\begin{algorithm}\label{alg:StaticCannon}
\caption{\name outer 2D algorithm}
\label{alg:StaticCannon}\footnotesize
\begin{algorithmic}[1]
\REQUIRE Inputs $A$, $B$, partitioned with $P$ processes and LI of size $\lambda$,
         process coordinates $(i,j,k) \in \tridentgrid$
\STATE $\C_{ijk} \gets \emptyset$
\STATE $h \gets (i+j) \bmod \sqrt{P/\lambda}$ \COMMENT{stagger step}
\FOR{$r$ \textbf{in} $\{0 \dots \sqrt{P/\lambda}\}$}
    \STATE $recvA \gets \text{Recv}(\A_{irk})$ \COMMENT{node-to-node communication $\A$} \label{line:nodecommA}
    \STATE $recvB \gets \text{Recv}(\B_{rjk})$ \COMMENT{node-to-node communication $\B$} \label{line:nodecommB}
    \STATE $\C_{ijk} \mathrel{+}= \operatorname{computePartialResult}(recvA, recvB)$ \label{line:computePartialResultcall}
\ENDFOR
\end{algorithmic}
\end{algorithm}

In \name, we integrate a Cannon-like stagger phase of $i + j$ \cite{cannon1969cellular}. 
In iteration $r$, each process $P_{ij}$ receives the same tiles as it would in the standard Cannon algorithm. 
The stagger step refers to an initial circular shift of submatrices on the 2D process grid that aligns matching submatrices for the first local multiplication. 
Blocks of $\A$ are shifted left by the process-row index, and blocks of $\B$ are shifted up by the process-column index, enabling subsequent iterations to rely solely on nearest-neighbor communication. 
In each iteration $r$, processes compute a partial product:
\begin{equation}\label{eq:tridaccess}
\C_{ij}^{(r)} = \A_{i(r+i+j)} \cdot \B_{(r+i+j)j},
\end{equation}

However, unlike the classic Cannon algorithm, our tiles are statically owned by the same process; therefore, we refer to this scheme as ``\textit{static} Cannon''.
\cref{alg:StaticCannon} provides the pseudocode for the static Cannon, which is the \name distributed SpGEMM algorithm applied to the 2D coarser-grained tiles.

\subsubsection{\name Inner 1D Algorithm}\label{sec:algorithm:inneralg}

Here, we describe how communication between coarser tiles is performed, how 1D slicing is integrated, and how the \name algorithm uses asynchrony.

\paragraph{Inter-node Communication}

As described in \cref{sec:algorithm:outeralg}, lines~\ref{line:nodecommA} and~\ref{line:nodecommB} in \cref{alg:StaticCannon} implement a nonstandard node-to-node communication pattern that we must define precisely. 
In our \tridentpart, 2D tiles $\A_{ir}$ and $\B_{rj}$ are not owned by a single process; instead, they are partitioned into 1D slices distributed across the $\lambda$ processes within the same fast \li group (i.e., within a single node). 
Communication between processes on different nodes must pass through the slower global interconnect \gi. 
To perform inter-node communication between nodes $N_{ir}$ and $N_{ij}$ (and vice versa), each process $P_{ir:}$, where ``:'' denotes the third inner index, in $N_{ir}$ is paired with a corresponding process $P_{ij:}$ in $N_{ij}$, and the relevant tile from, for example, $P_{irk}$ is sent to $P_{ijk}$.

\begin{figure}
    \centering
    \includegraphics[width=0.85\linewidth]{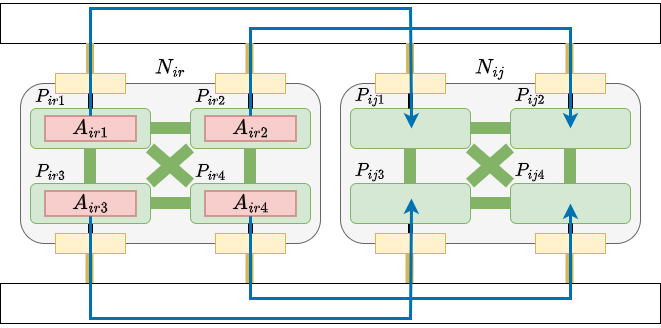}
    \caption{An example of internode communication between $N_{ir}$ and $N_{ij}$ for $\lambda = 4$. 
    The 2D tile $\A_{ir}$ is distributed across the $\lambda$ processes $P_{ir:}$. 
    During the internode communication, each process $P_{irk}$ sends its own tile to $P_{ijk}$ through \gi.} 
    \label{fig:outnode-dataflow}
\end{figure}

\cref{fig:outnode-dataflow} illustrates inter-node communication over a system with $\lambda = 4$.
This approach ensures that, after a single step of inter-node communication, the entire tile $\A_{ir}$ or $\B_{rj}$ is available on node $N_{ij}$. 
Then, subsequent redistribution within the node can proceed over the fast local interconnect without using the global interconnect.

\paragraph{Local Aggregation}\label{sec:algorithm:inneralg:localaggregation}

The data for the requested tile $\A_{ir}$ or $\B_{rj}$ are now available on node $N_{ij}$. 
However, these data are distributed across the $\lambda$ processes $P_{ij:}$ in node $N_{ij}$.
Each element of $\C_{ijk}^{(r)}$ is the scalar product of a row from the 2D tile $\A_{ir}$ and a column from the 2D tile $\B_{rj}$.
To compute $\C_{ijk}^{(r)}$, each process $P_{ijk}$ requires complete rows of $\A_{ir}$ and complete columns of $\B_{rj}$.
Due to the intranode 1D partitioning, each row of $\A_{ir}$ is stored entirely within the same process $P_{ijk}$, but columns of $\B_{rj}$ are distributed across the $P_{ij:}$ processes in $N_{ij}$. 
Thus, after inter-node communication, each $P_{ijk}$ requires the other 1D slices of $\B_{rj}$ held by the other $P_{ij:}$ processes on the same node $N_{ij}$.
Processes $P_{ij:}$ therefore perform an \allgatherv to reconstruct $\B_{rj}$ and compute the partial update $\C_{ijk} += \A_{irk} \cdot \B_{rj}$ on the fast local interconnect, independently within each node.

\begin{figure}
    \centering
    \includegraphics[width=0.547\linewidth]{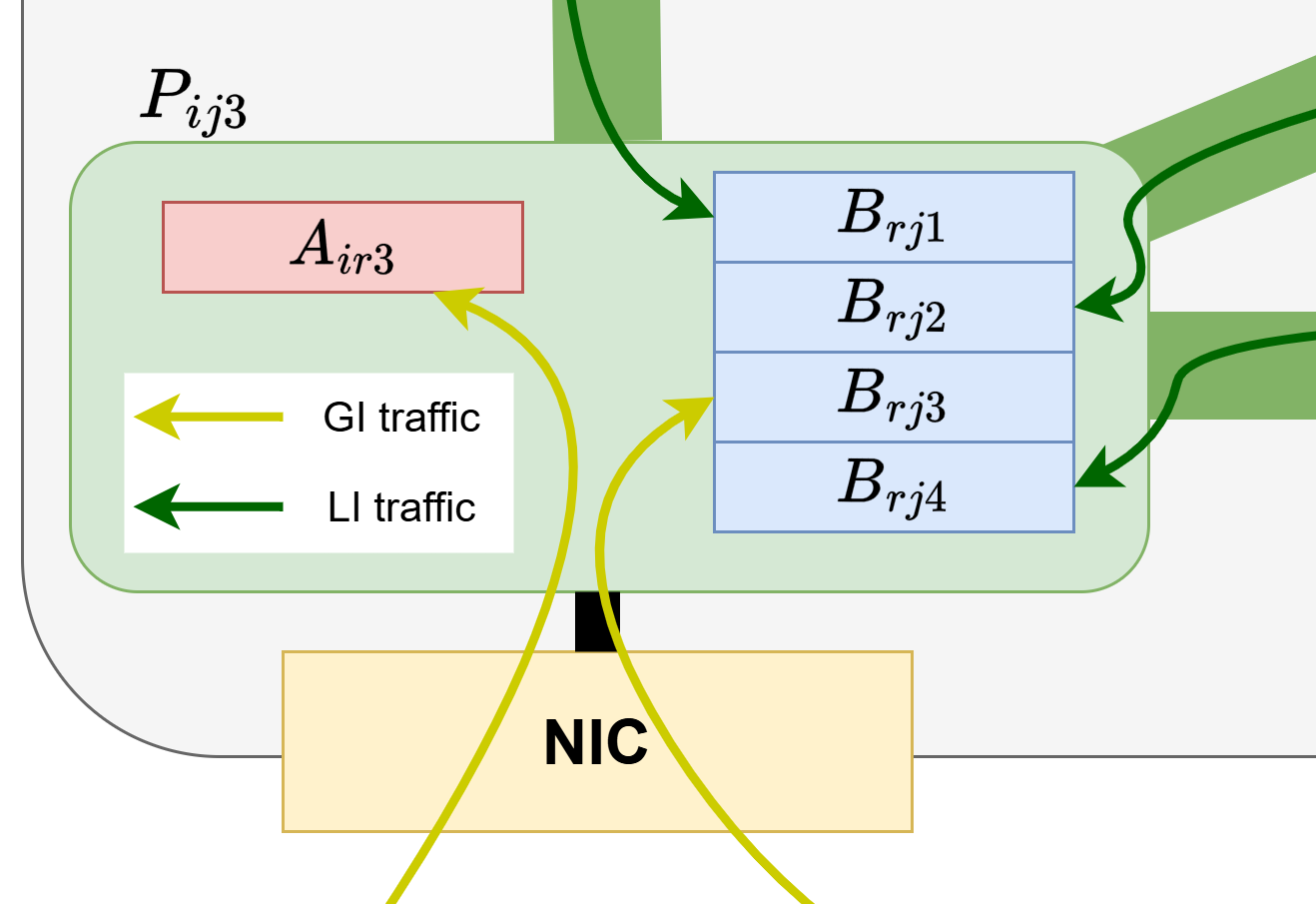}
    \caption{$P_{ij3}$ is a process associated with one of the 1D slices within a given 2D tile. 
    $P_{ij3}$ is mapped to one of the local GPUs within the node connected to other local GPUs via \li and to GPUs on other nodes via \gi. 
    In round $r$, $\A_{ir3}$ and $\B_{rj3}$ will come through \gi during node-to-node communication, while the other $\B_{rj:}$ will come through \li during the \allgatherv.
    } 
    \label{fig:dataflow}
\end{figure}

In \cref{fig:dataflow}, $P_{ij3}$ is a process associated with one of the 1D slices within a given 2D tile of the coarser-grained grid of size $\coarsergridcompact$, as shown in \cref{sfig:partitioning:our}. 
$P_{ij3}$ is mapped to one of the local GPUs within the node, which is connected to other local GPUs via \li and to GPUs on other nodes via \gi.
The figure illustrates the data flow for process $P_{ij3}$ at round $r$, where $\A_{rj3}$ and $\B_{rj3}$ come from the previous node-to-node communication through the global interconnect, and the other $\B_{rj:}$ tiles come from the \allgatherv operation through the local interconnect.
\cref{alg:base} expands line \ref{line:computePartialResultcall} of \cref{alg:StaticCannon}, detailing how intranode processes use the local interconnect and received tiles to compute the partial result. 

\begin{algorithm}\label{alg:base}
\caption{\name inner 1D algorithm}
\label{alg:base}\footnotesize
\begin{algorithmic}[1]
\REQUIRE Tiles $\A_{irk}$ and $\B_{rjk}$ received through an internode communication,
         process coordinates $(i,j,k) \in \tridentgrid$
\STATE $\B_{rj} \gets \text{AllGather}(\B_{rjk})$ \COMMENT{intra-node aggregation} \label{line:intranodecoll}
\STATE $\C_{ijk}^{(r)} = \A_{irk} \cdot \B_{rj}$
\RETURN $\C_{ijk}^{(r)}$
\end{algorithmic}
\end{algorithm}

\paragraph{Asynchronous Behavior}

\name uses a $\C$-stationary asynchronous \distspgemmalgo, where each tile can progress independently without global synchronization. 
In \name, asynchrony enables each node $N_{ij}$ to initiate communication independently, avoiding global synchronization and reducing idle time. 
Processes must wait for required tiles before computation but can simultaneously compute on available tiles and issue future tile requests.
This decoupling reduces idle time caused by uneven progress among processes.
Our local aggregation relies on an \allgatherv collective, making \name asynchronous at the coarse 2D tile level but synchronous within each node. 
Because the number of local GPUs per node, $\lambda$, is a small constant, this intranode synchronization is negligible as the total process count $P$ increases.

\paragraph{Communication Analysis}


The communication volume of \name can be characterized by assuming a uniform nonzero distribution typical of unstructured matrices and separating internode and intranode contributions to quantify the benefit of trident partitioning compared to traditional 2D approaches.

\begin{proposition}[Per-process communication volume]\label{prop:comm}
For unstructured matrices with $\sfrac{nnz}{P}$ nonzeros per tile, each iteration requires processor $P_{ij}$ to fetch tiles $\A_{ir}$ and $\B_{rj}$ from two remote processes, each transferring $\sfrac{nnz}{P}$ nonzeros over the global interconnect \gi. 
Then, including the intranode \allgatherv to reconstruct $\B_{rj}$, the per-iteration volume per process is
$\underbrace{2\cdot \sfrac{nnz}{P}}_{\gi}
+ \underbrace{(\lambda{-}1) \cdot \sfrac{nnz}{P}}_{\li}
= (\lambda{+}1)\cdot\sfrac{nnz}{P}$, where \li is the local interconnect. 
Over $\sqrt{\sfrac{P}{\lambda}}$ iterations, the internode volume per node is $\sfrac{2nnz}{(\sqrt{P}\sqrt{\lambda})}$.
This is a factor of $\sqrt{\lambda}$ less than Sparse SUMMA's $\sfrac{nnz}{\sqrt{P}}$ per process, with the remaining traffic absorbed by the local interconnect.
Each node performs these $\sqrt{\sfrac{P}{\lambda}}$ steps sequentially, while different nodes proceed asynchronously with respect to one another.
\end{proposition}
\newcommand{\Bp}{$\hat{B}_p$\xspace}
\newcommand{\Ap}{$\hat{A}_p$\xspace}

\section{Implementation}\label{sec:implementation}
 
\name uses hybrid distributed and shared memory parallelism, enabling asynchronous \distspgemmalgo at the outer 2D tier and fast collectives at the inner 1D tier. \name is available at \url{https://github.com/HicrestLaboratory/Trident}.

A request-based approach enables asynchrony, where each \textit{request} is a lightweight message containing the rank of the source process. 
Process $P_{ijk}$ sends a request to $P_{irk}$ or $P_{rjk}$ when a tile is needed to compute $\C_{ijk}$ at round $r$.  
Requests are stored in RDMA-accessible \emph{request queues} that can be updated asynchronously; the owner retrieves the source index and initiates synchronous communication to send the requested tile.
Outgoing communication, which involves sending $\A_{ijk}$ and $\B_{ijk}$, overlaps with local computation that includes receiving $\A_{irk}$ and $\B_{rjk}$ and computing $\C_{ijk}$ in round $r$. 
This is implemented using three \stdthreads per process: a \emph{main} thread for incoming communication and local computation, and two threads for outgoing $\A$ and $\B$ tiles.
During local aggregation, processes $P_{ij:}$ within a node $N_{ij}$ synchronize via \allgatherv to reconstruct the 2D tile $\B_{rj}$, with synchronization overhead negligible for large $P$ since $\lambda$ is a small constant, typically 4 or 8.

The remainder of this section first describes how \name uses one-sided asynchronous communication, then discusses implementation details of queues, threads, and intranode communication, and finally briefly describes the local SpGEMM.

\subsection{\name Asynchronous Algorithm}\label{sec:implementation:async}

In \name, internode asynchrony is implemented using MPI 3.0 one-sided primitives, which allow processes to expose memory accessible to others without synchronization or mutual rank knowledge.
This design is deliberate because mutual rank knowledge would require each process to track the state of others and could demand prohibitive memory to manage exchanges. 
Two communication threads are used: one handles outgoing $\A$ tiles, and the other handles outgoing $\B$ tiles. 
Each thread provides an RDMA-accessible request queue for asynchronous requests. 
The main thread issues send requests and computes local SpGEMM, while the communication threads continuously poll their queues and initiate communication when requests appear.
The request queue resides on the CPU, allowing the requesting process to issue a host-to-host \texttt{MPI\_Put} asynchronously without synchronization. 
The owner periodically checks the queue and triggers a device-to-device transfer via CUDA-aware MPI. 
Once the \texttt{MPI\_Put} completes, both processes perform the final transfer using synchronous peer-to-peer primitives.

\subsection{Internode Asynchronous Communication}\label{sec:implementation:communication:internode}

\begin{figure}
    \centering
    \includegraphics[width=\linewidth]{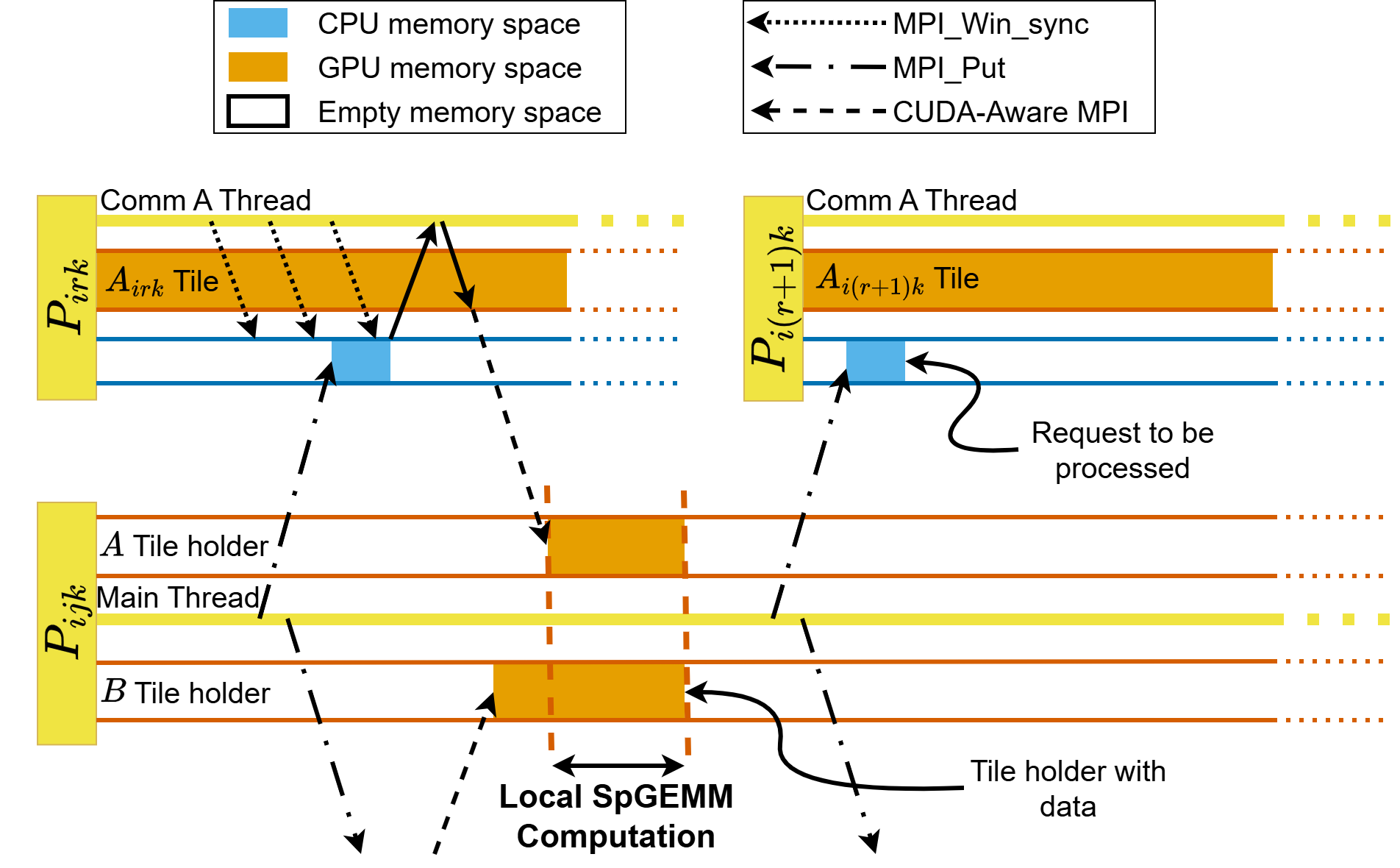}
    \caption{A snapshot of the three-thread model and asynchronous internode communication in a single SpGEMM round, where the main thread of $P_{ijk}$ requests $\A$ tiles using MPI\_Put and the communication thread provides them through CUDA-aware MPI.}
    \label{fig:multithreads}
\end{figure}

Each process $P_{ijk}$ uses the three-thread execution model introduced earlier in which threads operate independently and in parallel.
\Cref{fig:multithreads} illustrates a snapshot of the internode communication implemented in \name. 
The main thread of $P_{ijk}$ (shown as a thick line) issues requests for $\A$ tiles to the communication threads of $P_{irk}$ and $P_{i(r+1)k}$ (thick lines). 
For clarity, the figure omits the communication threads of $P_{ijk}$ and the main thread and threads of $P_{irk}$ and $P_{i(r+1)k}$ associated with $\B$ tiles.

The space between the orange thin lines indicates GPU memory spaces assigned to each thread. A solid region indicates valid data, while an empty region contains data that may be overwritten. 
The communication threads of processes $P_{irk}$ and $P_{i(r+1)k}$ maintain persistent GPU buffers that store their local input tiles of $\A$ and $\B$ (only $\A$ is shown) and remain unchanged throughout execution. 
The main thread of $P_{ijk}$ maintains a \emph{tile holder}, a persistent GPU buffer that is temporarily filled with received tiles and overwritten after each local SpGEMM, so copies of remotely received tiles exist only during the corresponding local computation.
 
The figure illustrates the dynamic interaction among processes during a single SpGEMM round. 
Regions between blue lines represent CPU memory spaces used to store request queues, which are associated with communication threads; the main thread does not maintain a request queue. 
An empty blue region indicates an empty queue, while a solid region indicates a pending request.
During round $r$, the $\A$ communication thread of $P_{irk}$ polls its request queue repeatedly and remains idle while the queue is empty, as indicated by the dotted lines.
Then, when the main thread of $P_{ijk}$ enqueues a request to $P_{irk}$ for an $\A$ tile (indicated by the upward dash-dot arrow) and subsequently issues a request for the corresponding $\B$ tile (downward dash-dot arrow), the $\A$ communication thread of $P_{irk}$ observes the now nonempty queue, processes the request, and sends its local $\A$ tile to $P_{ijk}$ (dashed arrow).
In parallel, the main thread of $P_{ijk}$ receives the $\B$ tile and begins the local SpGEMM as soon as the required tiles are available. 
Once the computation is complete, the main thread issues new requests to other processes, initiating the next SpGEMM round $r+1$.
Once the required internode and intranode tiles have been received, $P_{ijk}$ computes the aggregated tile $\B_{rj}$ (\cref{fig:dataflow}) and the partial result of $\C$, storing both in local persistent buffers not shown in \cref{fig:multithreads} because they remain within the process and are not involved in internode communication.

The three-thread design confines synchronization to the communication thread of $P_{irk}$ and the main thread of $P_{ijk}$, allowing the main thread of $P_{irk}$ to progress. 
The dotted and dash-dot lines highlight asynchronous one-sided communication.
Request queues are populated dynamically at runtime, depending on each process's progress.
Finally, for $P_{ijk}$, rounds with $r = j$ or $r = i$ skip intranode communication and use a local \cudamemcpy because $A_{irk}$ and $B_{rjk}$ are already in memory, avoiding queue and thread communication.

\subsection{Intranode Synchronous Communication}\label{sec:implementation:communication:intranode}

Once $P_{ijk}$ completes internode communication, the received $\B_{rjk}$ tile must be aggregated using an \allgatherv. 
A barrier is required to ensure that the processes $P_{ij:}$ in node $N_{ij}$ complete the operation before performing the local SpGEMM. 
Therefore, the \allgatherv is performed on the main threads.
This synchronization does not affect communication threads, which can continue to check their request queue and perform outgoing communication.

Based on measurements on Perlmutter, NCCL \allgatherv is faster than MPI CUDA-aware \allgatherv, which often relies on host staging. 
NCCL \allgatherv is not a dedicated call. 
It is implemented using $\mathsf{ncclSend}$ and $\mathsf{ncclRecv}$ operations wrapped in $\mathsf{ncclGroupStart}$ and $\mathsf{ncclGroupEnd}$, which automatically optimize communication.

The proposed trident partitioning is network-agnostic: it can be applied to any system that can be divided into groups where intra-group communication is faster than inter-group communication. For example, by replacing NCCL with RCCL, \name can also run on AMD machines.

\subsection{Local SpGEMM}\label{sec:impl:kokkos}

For local multiplication, we use \kokkos for portability~\cite{rajamanickam2021kk,DEVECI201833}. 
Its kernel uses hierarchical row-wise parallelism and a multi-level accumulator to balance load and exploit thread, team, and vector parallelism across heterogeneous execution cores, with architecture specific tuning of shared memory and team sizes. 
\kokkos is used for the local SpGEMM computation, and the results of each local SpGEMM are summed using \textsc{cuSPARSE}'s $\mathsf{spgeam}$~\cite{naumov2010cusparse}. 
\textsc{cuSPARSE}'s approach enables reuse of the workspace buffer for the sum, whereas \kokkos does not.

\section{Results}\label{sec:results}

This section evaluates \name, comparing it with several distributed SpGEMM implementations that support GPUs.
A runtime breakdown is also provided. 
Finally, we demonstrate \name's benefit by implementing a GPU-accelerated distributed version of Markov Clustering, a graph clustering algorithm that uses SpGEMM for random walks on the input graph \cite{hipmcl, van2008graph}.

\subsection{Baselines Description}

This section describes the SpGEMM baselines, including a state-of-the-art 1D algorithm and several 2D algorithms that are not hierarchy-aware, to highlight the benefits of trident partitioning.

\subsubsection{Trilinos}
The \textsc{Tpetra} package in Trilinos provides a GPU-accelerated 1D row-wise distributed SpGEMM algorithm~\cite{trilinos}.
Communication is organized as a series of nonblocking MPI\_Isend and MPI\_Irecv, where each process sends the specific rows of its local partition of $\B$ needed by other processes, based on the sparsity structures of their local partitions of $\A$. 
This communication is sparsity-aware, as the sparsity structure of $\A$ determines which rows of $\B$ are sent to each process. 
\textsc{Tpetra} uses \textsc{Kokkos} to perform the local SpGEMM. 
This baseline highlights \name's advantage over a sparsity-aware 1D SpGEMM for unstructured matrices.

\subsubsection{CombBLAS Sparse SUMMA}
Recent work by McFarland et al. implemented a GPU-accelerated version of Sparse SUMMA within the CombBLAS library~\cite{mcfarland2025parallel}. 
This 2D approach performs communication through a sequence of $\mathcal{O}(\sqrt{P})$ MPI\_Bcast. 
The communication is sparsity-oblivious. 
This baseline serves as a comparison to the dominant 2D hierarchy-oblivious distribution.

\subsubsection{Improved Sparse SUMMA}
In practice, the CombBLAS implementation is limited by costly host operations and frequent host-device data transfers, such as copying inputs and outputs and merging local SpGEMM results on the host.
As a result, its performance is generally worse than Trilinos, making it an uninformative baseline for unstructured matrices.
To provide a more fair comparison, we implemented an ``Improved Sparse SUMMA'' baseline that stores matrices on the device and merges intermediate results on the device, using the same data structures and libraries as \name. 
Experiments show that it is consistently faster than CombBLAS Sparse SUMMA, making it the main 2D baseline.

\begin{figure*}[t]
    \centering
    \includegraphics[width=\linewidth]{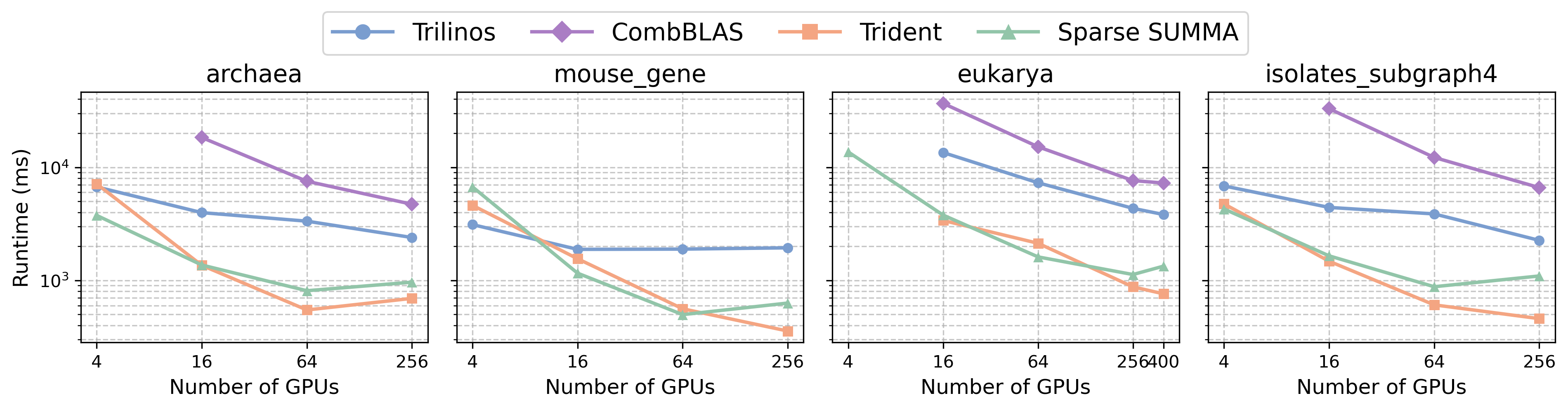}
    \caption{\name and baselines for strong scaling when squaring test matrices; missing data points indicate the implementation did not run successfully.}
    \label{fig:strong}
\end{figure*}

%
%

\subsection{Experimental Setup}
This section provides details about the datasets and hardware used to test the performance of \name.

\subsubsection{Hardware and Software Environment}

Our experiments were run on the GPU nodes of the NERSC Perlmutter supercomputer~\cite{perlmutter_nersc}. 
Each node contains 4 NVIDIA A100 GPUs, fully connected with 4 bidirectional NVLink 3.0 links between each pair, and a 64-core AMD EPYC 7763 CPU connected to each GPU via PCIe. 
The nodes are connected via a 3-hop dragonfly topology using the Slingshot 11 interconnect, with 4 Cassini NICs per node. 
Codes were compiled with NVCC 12.9 and g++ 13.2 using the \texttt{-O3} flag. 
The internode communication uses Cray MPICH 8.30 with GPU-aware support, while intranode communication uses NCCL 2.26. Local computation is handled with \kokkos 4.7.01 and \textsc{cuSPARSE} 12.9.

\begin{table}[t]
\centering
\caption{Density { ($\sfrac{nnz}{nrows \cdot ncols}$)} and size of matrices used in the evaluation. Imbalance is the ratio of max to average nnz per GPU at 256 GPUs.}
\label{tab:matrices}
\resizebox{.78\columnwidth}{!}{%
\begin{tabular}{lrrrc}
\toprule
\textbf{ID} & \textbf{Rows(=Columns)} & \textbf{Nnz} &  \textbf{Density} & \textbf{Imbalance} \\
\midrule
HV15R                & 2,017,169 & {283,073,458} & {6.95e-5} & 15.44 \\ 
mouse\_gene          &    45,101 & { 28,967,291} & {  0.014} & 4.65 \\ 
archaea              & 1,644,227 & {204,792,654} & {7.58e-5} & 1.02 \\ 
eukarya              & 3,243,106 & {359,763,936} & {3.42e-5} & 1.03 \\ 
isolates\_subgraph4  & 4,372,771 & {264,799,194} & {1.38e-5} & 1.02 \\ 
isolates\_subgraph5  & 2,186,385 & {66,399,522} & {1.38e-5} & 1.02 \\ 
cage15  & 5,154,859 & {99,199,551} & {3.73e-6} & 20.62 \\ 
uniparc & 2,856,197 & {29,621,573} & {3.63e-6} & 2.46 \\ 
reddit & 232,965 & {57,307,946} & {1.08e-3} & 1.15 \\ 
dielFilterV3real & 1,102,824 & {89,306,020} & {7.34e-5} & 10.19 \\ 
\bottomrule
\end{tabular}
}
\end{table}

\subsubsection{Dataset}

Our evaluation used large, unstructured, sparse matrices from real-world sources, as shown in Table~\ref{tab:matrices}. The matrices are from SuiteSparse~\cite{suitesparse} and the HipMCL repository~\cite{hipmcl}.

\subsection{Strong Scaling --- Matrix Squaring}

This section evaluates the scaling of \name compared to the baselines. For each matrix in Table \ref{tab:matrices}, we compute $\C = \A\A$, scaling to 256 GPUs except for \textit{eukarya} and \textit{reddit}, which are scaled to 400 GPUs due to their higher computational cost. 
Figure \ref{fig:strong} shows strong scaling results for all implementations on the \textit{eukarya}, \textit{archaea}, \textit{isolates\_subgraph4}, and \textit{mouse\_gene} matrices.
In \Cref{fig:strong}, data for CombBLAS on the \textit{mouse\_gene} matrix is unavailable because GALATIC \cite{Lett_2021}, the local SpGEMM library used by CombBLAS, fails on this input. 
In each experiment, CombBLAS is consistently the slowest, followed by Trilinos, while improved Sparse SUMMA (labeled Sparse SUMMA) is the fastest baseline and thus serves as the primary comparison to \name. The reported numbers represent the average of six trials, excluding the first trial.

Overall, \name consistently outperforms Trilinos and CombBLAS across matrices and most GPU counts, achieving a peak speedup of \(5.95\times\) over Trilinos on the \textit{mouse\_gene} matrix at 256 GPUs, and a geomean speedup of \(2.96\times\) across matrices at 256 GPUs.
The improved Sparse SUMMA is also faster than Trilinos and CombBLAS in most cases. 
On all matrices, \name matches or exceeds the performance of the improved Sparse SUMMA. 
For $P = \{4, 16\}$, that is, at small scale, the two implementations have comparable performance.
On $P = 64$, \name is faster for the \textit{archaea} and \textit{isolates\_subgraph4} matrices and performs similarly for \textit{eukarya} and \textit{mouse\_gene}. 
Importantly, at the largest scale, $P = 256$, \name consistently outperforms the improved Sparse SUMMA, with a speedup of up to \(2.38\times\) and a geomean speedup of \(1.54\times\) across matrices.
As $P$ increases, the performance advantage of \name becomes more pronounced, reflecting its improved handling of internode communication with a large number of GPUs.
Trident's node-level asynchrony allows nodes to progress independently, preventing imbalance caused by synchronization. This design tolerates small to moderate imbalance, as seen with \textit{mouse\_gene}, while intranode replication of $\A$'s coarse-grained tile reduces local skew.

\begin{figure}[t]
    \centering
    \includegraphics[width=\linewidth]{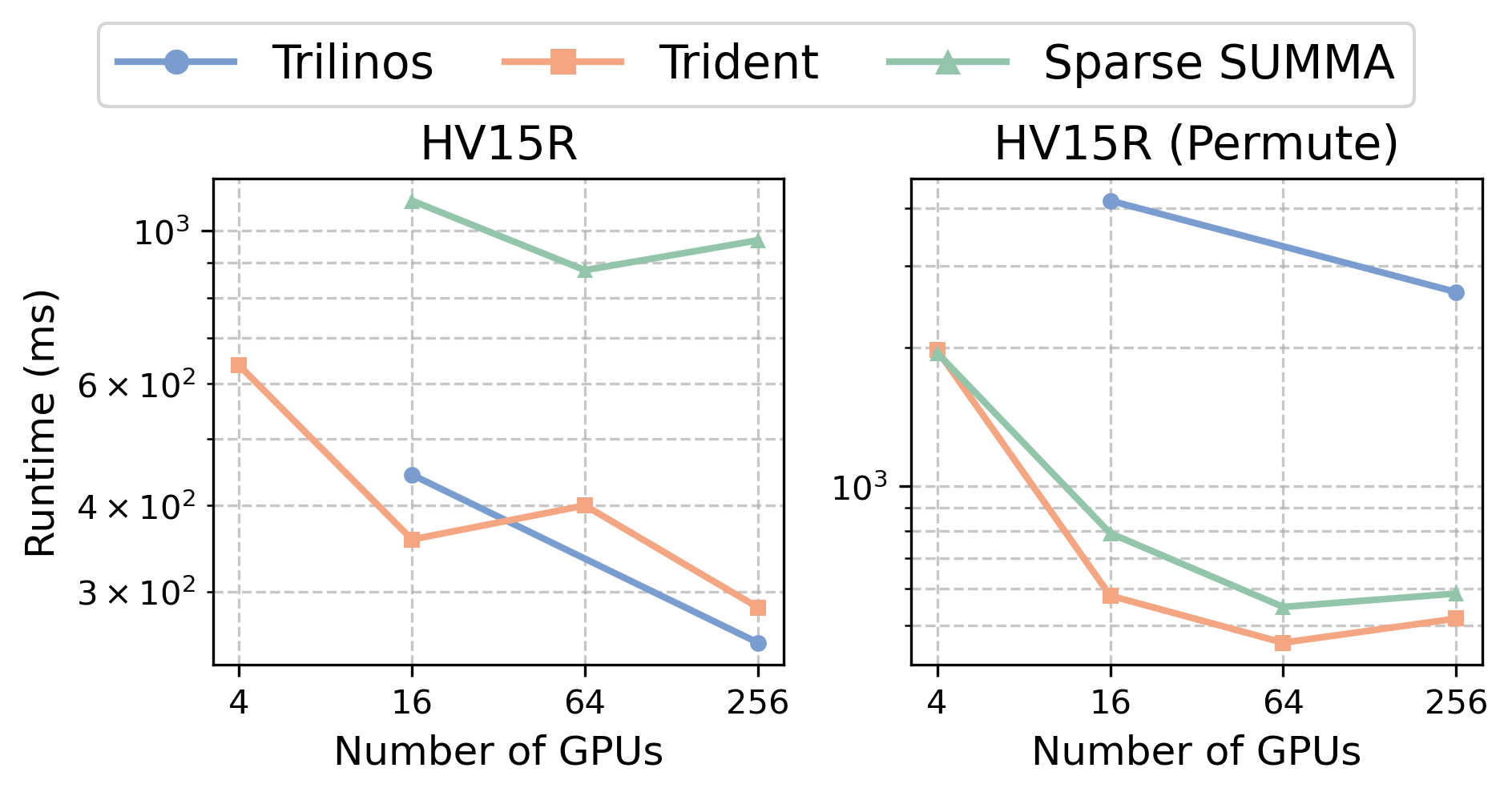}
    \caption{The strong scaling performance for \textit{HV15R}, with and without random permutation, is shown. The left plot shows strong scaling for each implementation without random permutation, while the right plot shows strong scaling after randomly permuting the rows and columns of $\A$; after permutation, \name is the fastest implementation.}
    \label{fig:permute}
\end{figure}

In Figure \ref{fig:permute}, we show that for structured matrices, a random permutation can be applied as lightweight preprocessing to take advantage of trident partitioning, and compare \name with baseline SpGEMM implementations on the structured matrix \textit{HV15R}, whose rows and columns are uniformly randomly permuted.
The \textit{HV15R} matrix contains several diagonal bands of nonzeros, making the unpermuted case highly structured. 
For the unpermuted matrix, Trilinos performs best, consistent with the effectiveness of 1D SpGEMM for structured sparsity. 
However, after permutation, which uniformly redistributes nonzeros and produces an unstructured matrix, both \summa and \name outperform Trilinos by up to $\approx 6\times$.
In unstructured matrices, permutations are unnecessary, while imposing structure on them, arguably more difficult than the reverse, would benefit 1D algorithms like Trilinos.
Results for CombBLAS are omitted here and in the following section because it was consistently slower than the other baselines.


\begin{figure*}[ht]
    \centering
    \includegraphics[width=\linewidth]{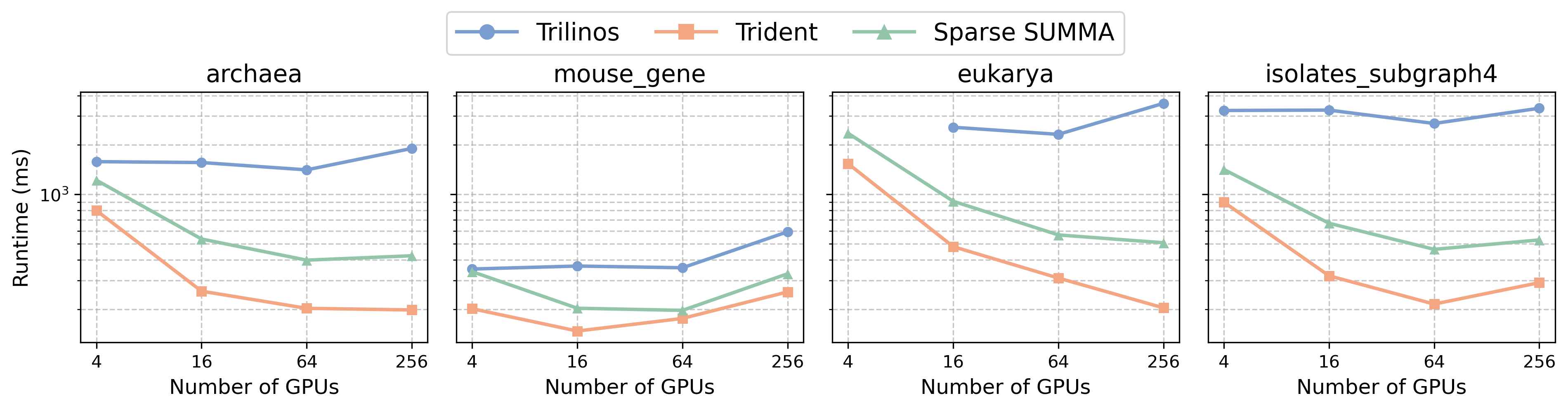}
    \caption{\name and baselines show strong scaling when multiplying each test matrix by a restriction operator; missing data points indicate the implementation did not run successfully.}
    \label{fig:restriction}
\end{figure*}

\subsection{Strong Scaling --- Restriction Operator}

This section evaluates Trident against the baselines when performing multiplication with a restriction operator $\mathbf{R}$, which is used in the setup phase of Algebraic Multigrid (AMG) \cite{dalton2015data, vanek1994algebraic}. 
Each experiment computes $\C = \A\mathbf{R}$, and the restriction operator is rectangular.

Figure \ref{fig:restriction} shows the strong scaling of each implementation when computing $\C = \A\mathbf{R}$ on GPU counts from 4 to 256.
\name is consistently the fastest implementation. 
On 256 GPUs, \name achieves a geomean speedup of $1.9\times$ over Improved Sparse SUMMA and $8.6\times$ over Trilinos.
On the \textit{archaea} and \textit{eukarya} matrices, \name scales up to 256 GPUs.
For \textit{mouse\_gene} and \textit{isolates\_subgraph4}, \name and other implementations do not scale to 256 GPUs. 
This is because computing $\A\mathbf{R}$ typically requires fewer FLOPS than computing $\A\A = \A^2$ due to the smaller number of nonzeros in $\mathbf{R}$, resulting in less computation to saturate each GPU.
Overall, \name outperforms other implementations in each setting, demonstrating its effectiveness beyond $\A^2$.

\subsection{Runtime Breakdown}

\begin{figure}[t]
    \centering
    \begin{subfigure}[t]{0.4\textwidth}
        \centering
        \includegraphics[width=\linewidth]{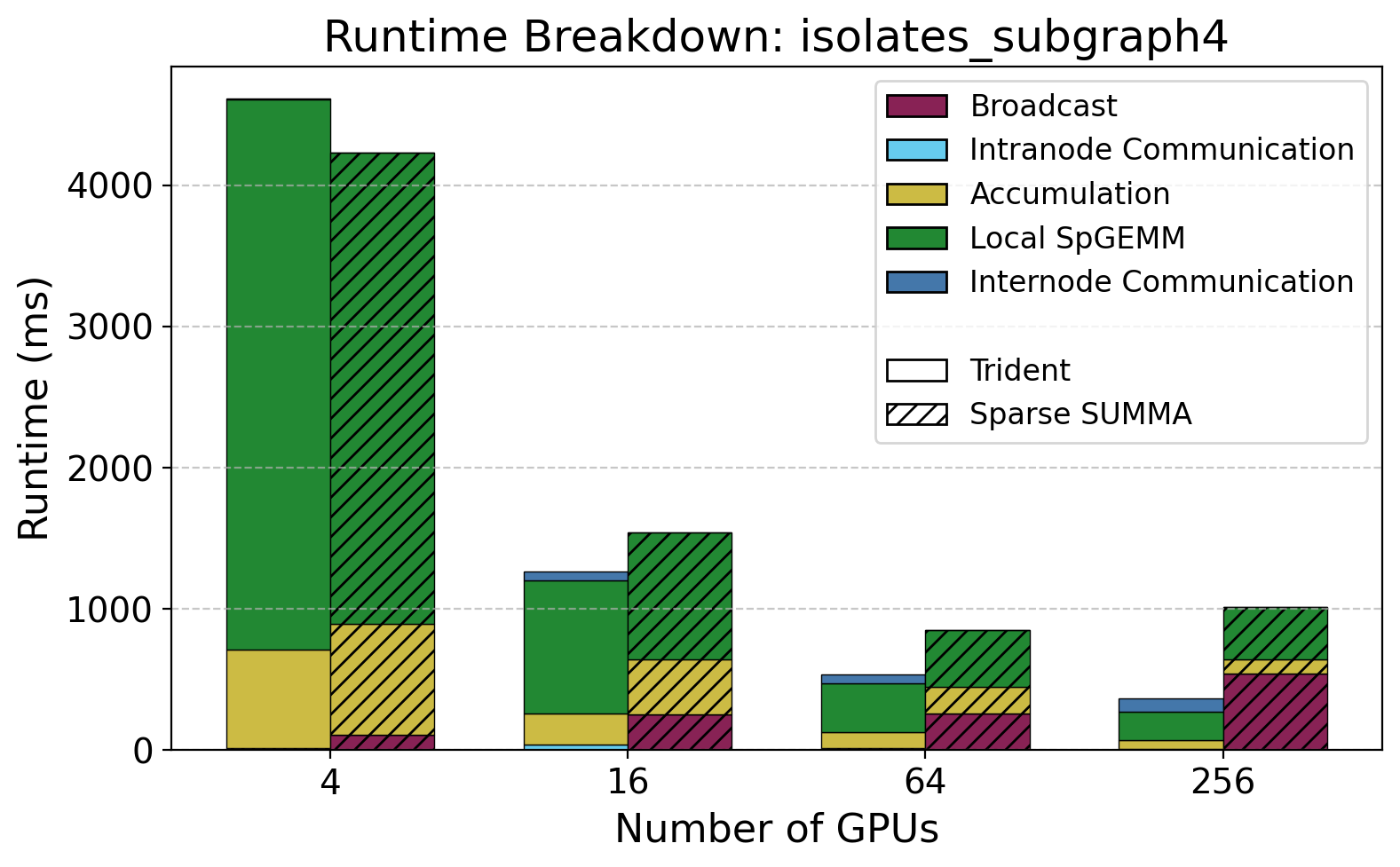}
    \end{subfigure}
    \hfill
    \begin{subfigure}[t]{0.4\textwidth}
        \centering
        \includegraphics[width=\linewidth]{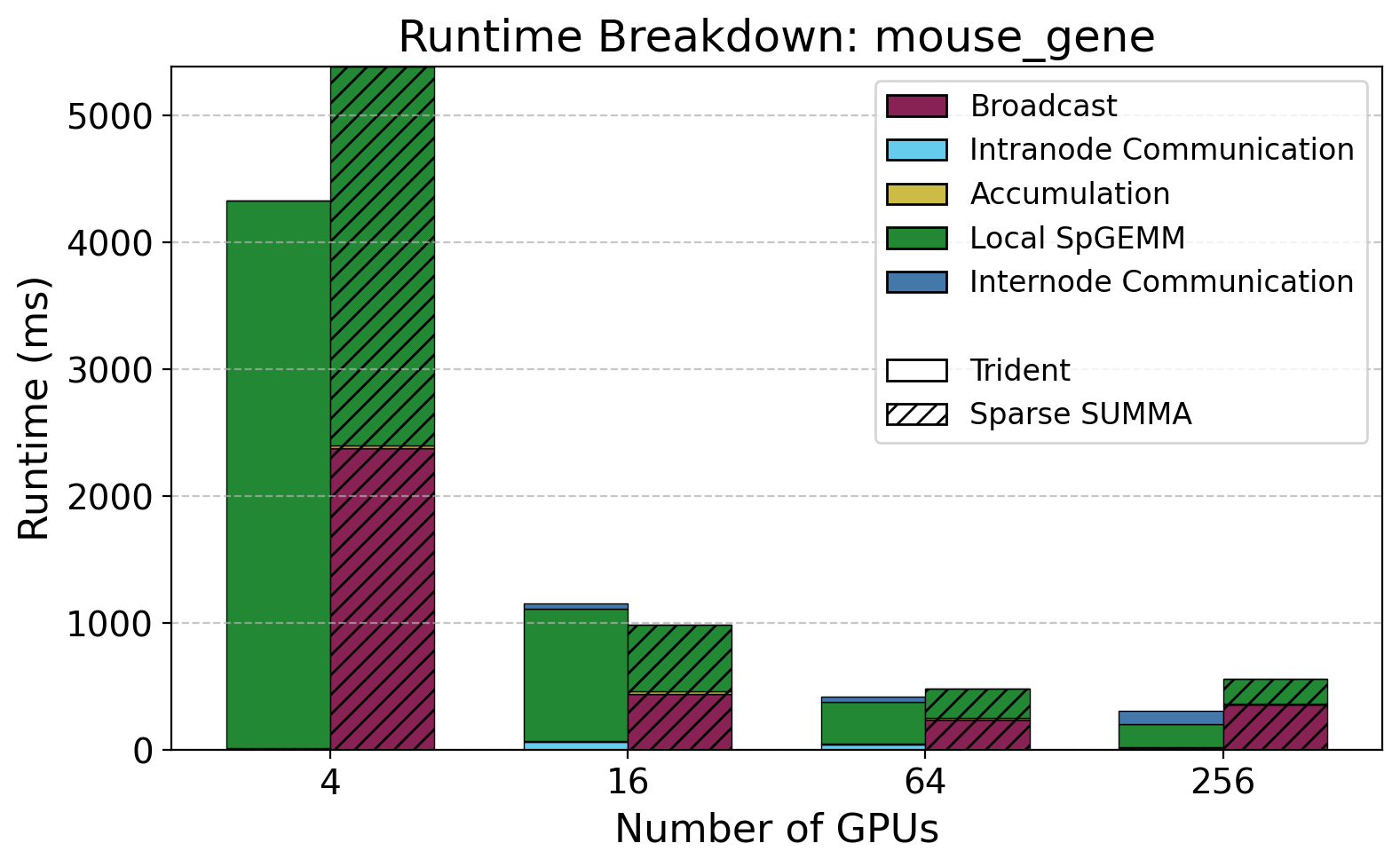}
    \end{subfigure}
    \caption{Runtime breakdown of \name and \summa on \textit{mouse\_gene} and \textit{isolates\_subgraph4}.}
    \label{fig:runtime-bdown}
\end{figure}

A runtime breakdown of \name and \summa illustrates how their algorithmic components scale. 
The comparison is limited to \summa, as it consistently outperformed the other baselines. 
Figure~\ref{fig:runtime-bdown} illustrates the time spent in each phase when squaring the \textit{isolates\_subgraph4} and \textit{mouse\_gene} matrices, which are representative of the full matrix set. 
Phase times are reported as averages across participating processes. 

The breakdown shows that communication in \name scales more favorably than in \summa, in accordance with our communication analysis in \Cref{prop:comm}.
The intranode communication in \name is negligible due to the high intranode bandwidth supporting the \allgatherv. 
The internode communication increases slightly with GPU count but remains a small fraction of the total runtime and scales better than the broadcast communication in \summa. 
Consequently, the aggregate communication cost of \name is lower than that of \summa, with the gap widening at larger scales, which explains the improved scalability of \name.

\subsection{Communication Volume}

\begin{figure}
    \centering
    \includegraphics[width=\linewidth]{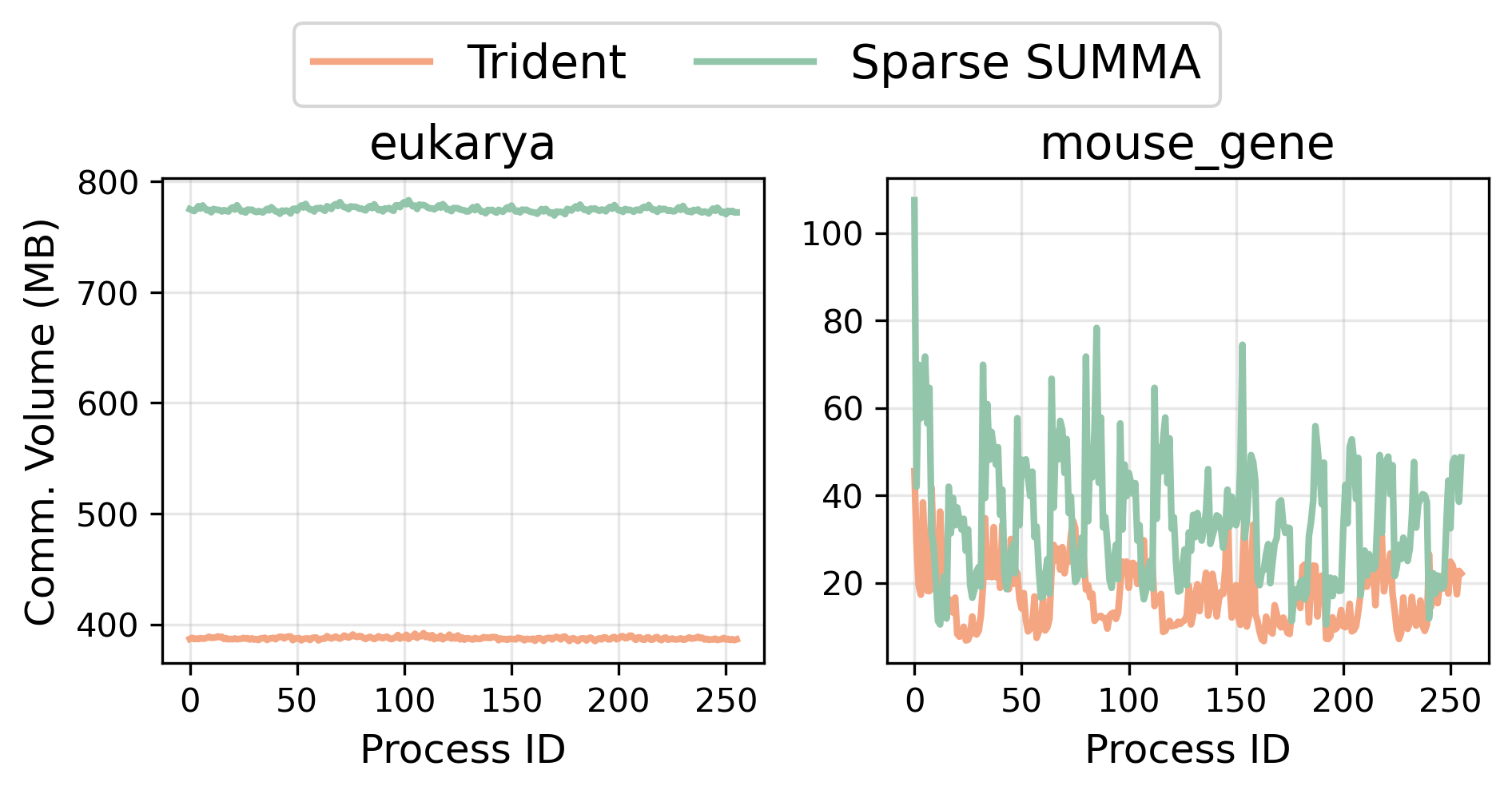}
    \caption{The internode communication volume of \name compared to \summa.}
    \label{fig:vol}
\end{figure}

The reduction in internode communication volume achieved by \name relative to \summa is illustrated in Figure~\ref{fig:vol}, which reports the total data sent by each process over the global interconnect \gi.
For the \textit{mouse\_gene} matrix, per-process communication volume varies across GPUs, but \summa generally sends more data than \name.  
In a small number of processes, \name can send slightly more data because the nonzeros in the tiles of the input matrices can increase when the dimensions of the local submatrices in \name partitioning differ from those in a standard 2D partitioning, which could theoretically lead to tiles with many more nonzeros. 
This effect is limited, and as shown in Figure~\ref{fig:runtime-bdown}, the overall communication time for \textit{mouse\_gene} remains lower for \name.
For the \textit{isolates\_subgraph4} matrix, the more uniform nonzero distribution leads to a reduction in communication volume across processes by $\approx 2\times$.
Overall, \name reduces internode communication volume relative to \summa for both matrices.

\subsection{Markov Clustering}

Finally, we evaluate the benefits of \name in the context of a real application that relies on SpGEMM: Markov Clustering (MCL)~\cite{van2008graph}. 
MCL is a widely used graph clustering algorithm based on random walks that identifies tightly connected subgraphs.
Each iteration consists of squaring a sparse matrix using SpGEMM (the expansion step), normalizing the columns, pruning entries below a user-defined threshold, and applying an element-wise square to the remaining entries.
MCL typically operates on unstructured matrices from domains such as computational biology~\cite{hipmcl}, and repeated squaring and pruning further eliminate any remaining structure.
These characteristics make MCL a representative and well-suited application for evaluating \name.

\begin{figure}[t]
    \centering
    \includegraphics[width=\linewidth]{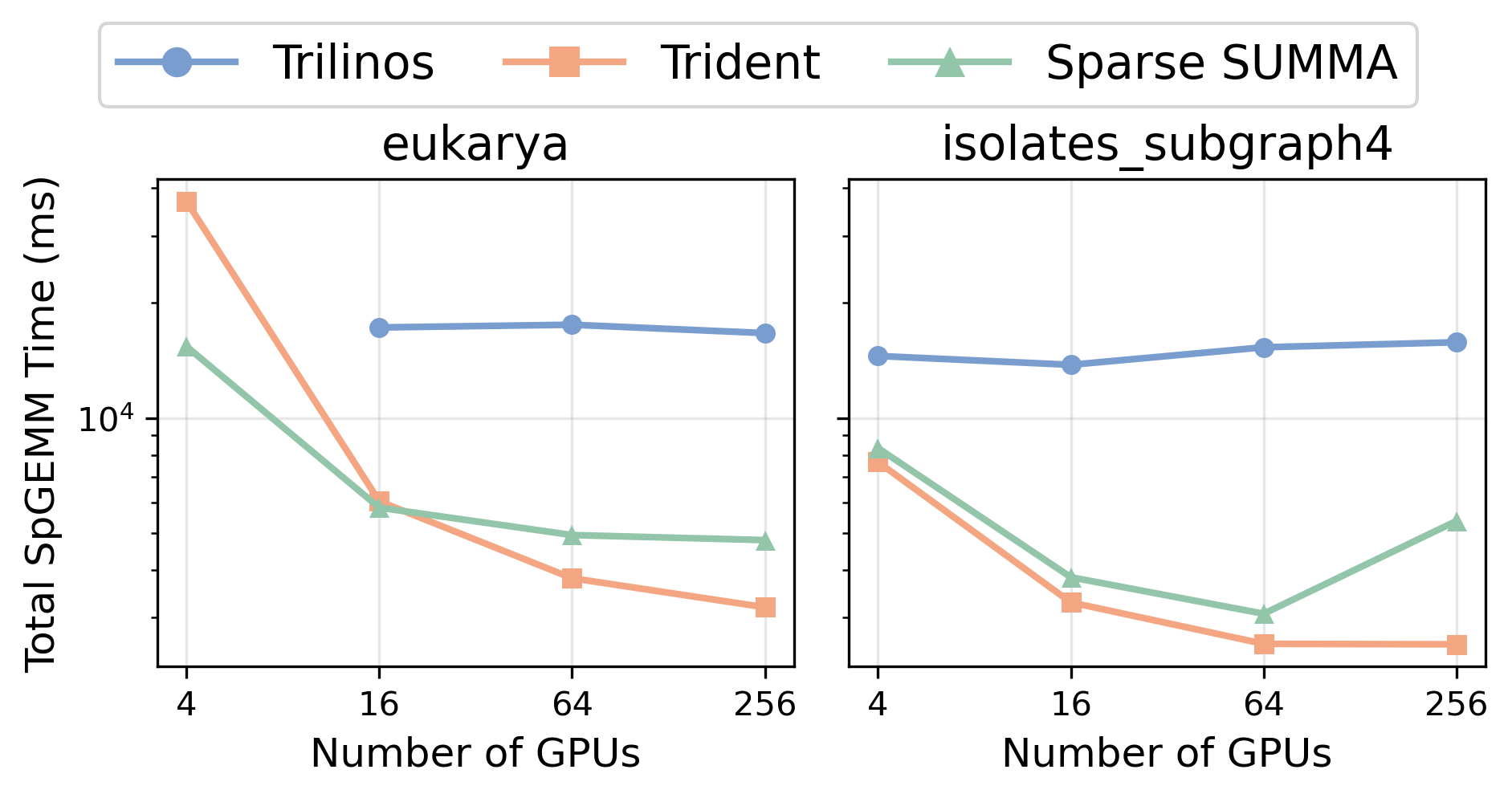}
    \caption{The scaling performance of MCL expansion step.}
    \label{fig:mcl}
\end{figure}

Three versions of MCL were implemented, differing only in the expansion step: one using \name, one using Trilinos, and one using \summa. 
Each version was run for 10 iterations with a pruning threshold of 0.002, and the time spent in the expansion step was compared. 
Figure~\ref{fig:mcl} shows the strong scaling of the expansion step for the \textit{eukarya} and \textit{isolates\_subgraph4} matrices, which were chosen because they represent large, real-world protein similarity graphs, a common application of MCL~\cite{hipmcl}.

Both \name and \summa consistently outperform Trilinos, with \name achieving up to a $t\times$ speedup over \summa at higher GPU counts. 
Runtime decreases across iterations as the matrix becomes sparser while clustering proceeds.
The plot is omitted for space, but in early iterations, when the matrix contains many nonzeros, \name is noticeably faster than both Trilinos and \summa, especially at 64 and 256 GPUs.
The best speedup for \textit{isolates\_subgraph4} is at 256 GPUs, approximately $2.5\times$. 
Overall, these results highlight the benefits of \name for real-world distributed SpGEMM applications.

\section{Conclusion}

In this work, we presented \name, a hierarchy-aware distributed sparse matrix multiplication algorithm designed to exploit modern multi-GPU supercomputing network architectures. 
By exploiting high-bandwidth intranode interconnects and minimizing internode communication, \name achieves significant speedups over existing 1D and 2D approaches on real-world unstructured matrices.
The core contribution of \name is the \textbf{trident partitioning} scheme, which combines 2D partitioning across nodes with 1D partitioning within a node. This enables efficient pipelining of internode transfers with intranode collectives, reducing both latency and communication volume.
To the best of our knowledge, \name is the first hierarchy-aware distributed SpGEMM algorithm.

Our evaluation showed that \name is consistently faster than Trilinos and CombBLAS, with a peak speedup of approximately $\approx 5.95\times$ over Trilinos. 
Compared to the Improved Sparse SUMMA baseline, \name achieves comparable or better performance, with speedup reaching $\approx 2\times$ at a large number of GPUs. 
In addition, our runtime breakdown and communication volume analyses confirm that \name effectively reduces internode communication while fully utilizing intranode bandwidth. 
In future work, we will explore hierarchical implementations of other distributed sparse primitives, such as 3D and sparsity-aware 1D SpGEMM.

\begin{acks}
This research used resources from the National Energy Research Scientific Computing Center, a DOE Office of Science User Facility supported by the Office of Science of the U.S. Department of Energy under Contract No. DE-AC02-05CH11231, using NERSC award ASCR-ERCAP0030076. 
This material is based upon work supported by the U.S. Department of Energy, Office of Science, Office of Advanced Scientific Computing Research, Department of Energy Computational Science Graduate Fellowship under Award Number DE-SC0025528.
Partially Funded by the European Union. 
The views and opinions expressed are however those of the author(s) only and do not necessarily reflect those of the European Union or the European High-Performance Computing Joint Undertaking. 
Neither the European Union nor the granting authority can be held responsible for them. GA n. 101175702 ``Network for European Exascale Systems''.
This work was partially developed during Lorenzo Pichetti's research visit to Cornell. 
The authors disclose that generative AI and editing assistants were used to assist with grammar checking and to improve the clarity of the writing.
\end{acks}

\bibliographystyle{ACM-Reference-Format}
\bibliography{ref} 

\end{document}